\documentclass[aps,prl,twocolumn,showpacs]{revtex4}
\usepackage{graphicx}
\usepackage{bm}
\usepackage{amsmath}
\begin{document}
\title{Vortex State in
Na$_x$CoO$_2$$\cdot y$H$_2$O: $p_x\pm ip_y$-wave versus
$d_{x^2-y^2}\pm id_{xy}$-wave Pairing}
\author{Qiang Han$^{1,2}$}
\author{Z. D. Wang$^{1,2}$} %
\email{zwang@hkucc.hku.hk}%
\author{Qiang-Hua Wang$^{3}$}
\author{Tianlong Xia$^{1}$}
\address{$^1$Department of Physics, University of Hong
Kong, Pokfulam Road, Hong Kong, China } %
\address{$^2$Department of Material Science and Engineering, University
of Science and Technology of China, Hefei Anhui 230026,
China}%
\address{$^3$National Laboratory of Solid State Microstructures, Institute for Solid
State Physics, Nanjing University, Nanjing 210093, China}
\date{\today}

\begin{abstract}
Based on an effective Hamiltonian specified in the  triangular
lattice with possible $p_x\pm ip_y$- or $d_{x^2-y^2}\pm
id_{xy}$-wave pairing,
 which has close relevance to the newly discovered
Na$_{0.35}$CoO$_2$$\cdot y$H$_2$O, the electronic structure of the
vortex state  is studied by solving the Bogoliubov-de Gennes
equations. It is found that $p_x\pm ip_y$-wave is favored for the
electron doping as the hopping integral $t<0$. The lowest-lying
vortex bound states are found to have respectively zero and
positive energies for $p_x\pm ip_y$- and $d_{x^2-y^2}\pm
id_{xy}$-wave superconductors, whose vortex structures exhibit the
intriguing six-fold symmetry.
 In the presence of strong on-site repulsion, the antiferromagnetic and
ferromagnetic orders are induced around the vortex cores for the
former and the latter, respectively, both of which cause the
splitting of the LDOS peaks due to the lifting of spin degeneracy.
The microscopic STM and the spatially resolved NMR measurements are able
to probe the new features of
vortex states uncovered in this work.

\end{abstract}
\maketitle

The recent discovery of superconductivity in the Co oxide,
Na$_{0.35}$CoO$_2$$\cdot y$H$_2$O \cite{takada}, has intrigued
much interest on its novel properties especially the similarities
to and differences from the high-$T_c$ copper oxides.
Superconductivity occurs  after sodium content is reduced in
Na$_{0.75}$CoO$_2$ and the distance between the CoO$_2$ planes is
enlarged by hydration, indicating that the superconductivity is
mostly relevant to the two-dimensional CoO$_2$ layer similar to
the role of CuO$_2$ layers in cuprates. Furthermore, the Co$^{4+}$
atoms in neutral (undoped) CoO$_2$ plane has spin-$\frac{1}{2}$,
resulting in the parent compound a spin-$\frac{1}{2}$
antiferromagnet. On the other hand, because the spins form a
triangular lattice, the antiferromagnetism is frustrated and the
resonating-valence-bond (RVB) state \cite{anderson} might give
rise to superconductivity under proper doping. At present, the
mechanism of the superconductivity in this material is hotly
debated and accordingly the pairing symmetry of the
superconducting order parameter (OP) has been paid significant
attention although still controversial. Theories
\cite{baskaran,kumar,qhwang} based on the RVB theory support the
view that the superconducting OP has the spin-singlet
broken-time-reversal-symmetry (BTRS) chiral $d_{x^2-y^2}\pm
id_{xy}$ symmetry, while theories based on a combined symmetry
analysis with fermiology \cite{tanaka} and numerical calculations
\cite{singh,jni} of normal state electronic structure speculate
that the OP is spin-triplet BTRS chiral $p_x\pm ip_y$-wave
symmetry. The experimental results reported by different groups
using the same nuclear-magnetic-resonance(NMR) technique are also
confusing:  one group \cite{waki} supports the spin-triplet
$p_x+ip_y$-wave symmetry while another group \cite{kobayashi}
claimed to support the spin-singlet $s$-wave symmetry.

In this Letter, to have valuable clues for experimental
clarification, we elucidate and compare the effects of the two
most possible pairing symmetries,
$p_x+ip_y$\cite{tanaka,singh,jni} and $d_{x^2-y^2}+id_{xy}$
\cite{baskaran,kumar,qhwang}waves, on the electronic structure of
the vortex state. In particular, we shall answer two crucial
questions clearly: (i) what are the new features of the vortex
state in this kind of triangular system? (ii) what are the
experimentally observable signatures showing the differences
between the mentioned two pairing symmetries? Because the
mechanism of the superconductivity in Na$_{0.35}$CoO$_2$$\cdot
y$H$_2$O is still unclear at present, we will not adopt the
well-known $t$-$J$ model as in Refs. [\onlinecite{kumar,qhwang}]
as it only gives rise to the spin-singlet pairing. Here the
well-established $t$-$U$-$V$ Hubbard model \cite{jxzhu} is
employed with competing magnetic ($U$) and superconducting ($V$)
interactions. Although phenomenological, this model captures the rich
physics of system with competing orders and has been applied to
study the field-induced antiferromagnetic and charge-density-wave
(CDW) orderings around the vortex core of high-$T_c$ $d$-wave
cuprates\cite{tuvmodel}, having accounted for several important
experimental observations. Considering similarities of this new
superconductor to the cuprates as well as possibilities of
both  spin singlet and triplet
pairings, we extend this model to study the
superconducting cobalt oxide with either spin singlet or triplet
pairing channel in the triangular lattice and examine the novel
properties in the vortex state. The effective model Hamiltonian is
expressed as
\begin{eqnarray}
H_{\text{eff}}&=&-\sum_{\langle i,j\rangle \sigma
}(t_{ij}c_{i\sigma }^{\dagger }c_{j\sigma
}+\text{h.c.})+\sum_{i,\sigma }(Un_{i%
\bar{\sigma }}-\mu )c_{i\sigma }^{\dagger }c_{i\sigma } \nonumber\\%
&&+\sum_{\langle i,j\rangle} \left [\Delta
_{ij}^{\pm}(c_{i\uparrow }^{\dagger }c_{j\downarrow }^{\dagger}\pm
c_{i\downarrow }^{\dagger }c_{j\uparrow
}^{\dagger})+\text{h.c.}\right ],
\end{eqnarray}%
where $n_{i\sigma}=\langle
c_{i\sigma}^{\dagger}c_{i\sigma}\rangle$ is the electron density
with spin $\sigma$. $\mu$ is the chemical potential. $\pm$ is for
spin triplet and singlet pairings, respectively and the pairing
potential $\Delta_{ij}^{\pm}$ is defined as $
\Delta_{ij}^{\pm}=\frac{V}{2}(\langle
c_{i\uparrow}c_{j\downarrow}\rangle \pm \langle
c_{i\downarrow}c_{j\uparrow}\rangle), $ which comes from a mean
field treatment of the pairing interaction $V\sum_{\langle
i,j\rangle}(c_{i\uparrow }^{\dagger }c_{j\downarrow
}^{\dagger}c_{i\uparrow }c_{j\downarrow }+c_{i\downarrow
}^{\dagger }c_{j\uparrow }^{\dagger}c_{i\downarrow }c_{j\uparrow
})$. In an external magnetic field, the hopping integral $t_{ij}$
can be written as $t_{ij}=t\exp(i\varphi_{i,j})$ for the
nearest-neighbor (NN) sites $\langle i,j\rangle$, where
$\varphi_{i,j}=-\frac{\pi }{\Phi _{0}}\int_{{\bf{r}}_{i}}^{{\bf{r}}_{j}}%
{\bf{A}}({\bf{r}})\cdot d{\bf{r}}$ with ${\bf{A}}({\bf{r}})$ the
vector potential and $\Phi _{0}=hc/2e$ the superconducting flux
quantum. The internal field induced by supercurrents around the
vortex core is neglected since Na$_{0.35}$CoO$_2$$\cdot y$H$_2$O
can be treated as extreme type-II superconductors according to
experiment \cite{sakurai} estimation.
Therefore, $%
{\bf{A}}({\bf{r}})$ is approximated as $(0,Bx,0)$ in a Landau
gauge where $B$ is the external magnetic field. By applying the
self-consistent mean-field approximation and performing the
Bogoliubov transformation, diagonalization of the Hamiltonian
$H_{\text{eff}}$ can be achieved by solving the following
Bogoliubov-de Gennes (BdG) equations:
\begin{equation}
\sum_{j}\left(
\begin{array}{cc}
H_{ij,\sigma} & \Delta^{\pm}_{i,j} \\
\mp\Delta _{i,j}^{\pm\ast } & -H_{ij,\bar{\sigma}}^{\ast}
\end{array}%
\right) \left(
\begin{array}{c}
u_{j,\sigma}^{n} \\
v_{j,\bar{\sigma}}^{n}%
\end{array}%
\right) =E_{n}\left(
\begin{array}{c}
u_{j,\sigma}^{n} \\
v_{j,\bar{\sigma}}^{n}
\end{array}%
\right)  \label{BdG}
\end{equation}%
where $u^{n},v^{n}$ are the Bogoliubov quasiparticle amplitudes \
with corresponding eigenvalue $E_{n}$.
$H_{ij,\sigma}=-t_{ij}+\delta_{i,j}(Un_{i\bar{\sigma}}-\mu)$ with
$n_{i\sigma}$ subject to the self-consistent conditions:
$n_{i\uparrow}=\sum_n\{|u^n_{i,\uparrow}|^2f(E_n)$ and
$n_{i\downarrow}=\sum_n\{|v^n_{i,\downarrow}|^2[1-f(E_n)]$ with
$f(E)$ the Fermi distribution function. $\Delta_{i,j}$ is
calculated according to: $
\Delta^{\pm}_{i,j}=\frac{V}{4}\sum_{n}(u_{i\uparrow}^n
v_{j\downarrow}^{n*}\mp u_{j\uparrow}^n
v_{i\downarrow}^{n*})\tanh(\frac{E_n}{2k_B T}). %
\label{delta}
$

%For the lack of experimental data about the structure of vortex
%lattice in Na$_x$CoO$_2$$\cdot y$H$_2$O, we assume that the vortex
%lattice is triangular with perfect matching with the underlying
%CoO$_2$ plane.
In this work, we choose such a  magnetic unit cell (MUC), which
accommodates two superconducting flux quanta $2\Phi_0$\cite{ywang}
and is characterized by ${\bf{R}}_1=N{\bf{a}}_1$ and
${\bf{R}}_2=2N{\bf{a}}_2$ with $N$ an integer.
${\bf{a}}_1=a(\sqrt{3}/2,1/2)$ and ${\bf{a}}_2=a(0,1)$ are the
prime translation vectors of the CoO$_2$ triangular lattice. By
introducing the magnetic Bloch states labelled by the
quasimomentum, we can handle an array of MUC's under the modified
periodic boundary condition related to the phase factor by
magnetic translations $\chi({\bf{r}},{\bf{R}})=-2\pi
my-\pi(m-2n+m^2/2)$ with ${\bf{R}}=m{\bf{R}}_1+n{\bf{R}}_2$. In
addition, we set $t<0$ \cite{comment1} according to the analysis
on the band calculation
%Refs. [\onlinecite{baskaran,qhwang}]
\cite{baskaran,qhwang,singh}. The energy and length will be measured in
units of $|t|$ and the lattice constant $a$.

It is noteworthy that the coexistence of and competition between
superconductivity and ferromagnetism(FM)/antiferromagnetism(AFM)
in this novel material is  very interesting and important, but
quite complicated even in the homogenous case.
%  issue just as in the high-$T_c$ cuprates and
%heavy fermion compound UGe$_2$\cite{saxena}.
 We leave it for
future careful investigation. In the present work, we  focus only
on the vortex state, %
%and simply assume here that in the doping
%region we are interested in, both the three-sublattice long-range
%AFM and the itinerant FM orders caused by $U$ are weak and fully
%overwhelmed by the superconducting order at zero temperature,
%In other words, in homogenous superconducting systems at
%zero-temperature, the $U$ term is merely renormalizing the
%chemical potential without lifting the degeneracy of the spin
%degree of freedom
bearing in mind that magnetic orders may be induced in the vortex
cores where superconductivity is significantly suppressed. In the
homogenous superconducting state, we find that   the $p_x\pm
ip_y$-wave pairing state is always favored for $\bar{n}>1$ while
the $d_{x^2-y^2}\pm id_{xy}$-wave case is stablized for
$0.5<\bar{n}<$1\cite{comment1} with $\bar{n}$ the average electron
number per site.

\begin{figure}
\includegraphics[width=6.5 cm]{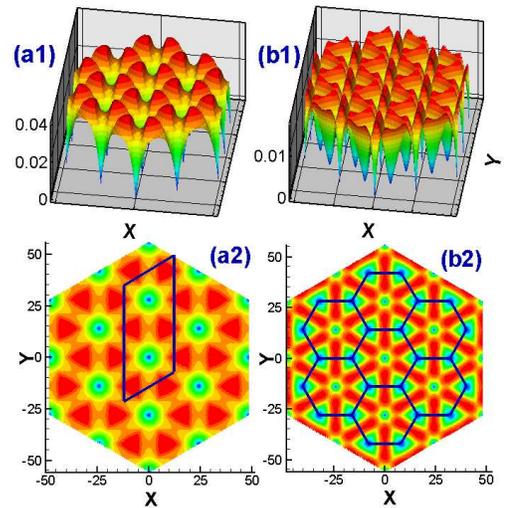}
\caption{\label{p-fll}3D and contour plots of the spatial
distribution of the dominant $|\Delta^{p_x+ip_y}|$ (a) and the
induced subdominant $ |\Delta^{p_x-ip_y}|$ (b). The blue
parallelogram in (a2) denotes the $28\times 56$ MUC in our study.
See text for detail.}
\end{figure}%
Now we address the vortex lattice structure of the gauge-invariant
$\Delta^{p_x\pm ip_y}$ and  $\Delta^{d_{x^2-y^2}\pm id_{xy}}$
according to $\Delta^{p_x\pm
ip_y}({\bf{r}}_i)=\sum_{{\bf{\delta}}}
\Delta^{+}_{i,i+{\bf{\delta}}}e^{\mp i\theta (\delta)}
e^{i\varphi_{i,i+\delta}}/6$ and $\Delta^{d_{x^2-y^2}\pm
id_{xy}}({\bf{r}}_i)=\sum_{{\bf{\delta}}}
\Delta^{-}_{i,i+{\bf{\delta}}}e^{\mp 2i\theta (\delta)}
e^{i\varphi_{i,i+\delta}}/6$, where $i+\delta$ denotes the six NN
sites of the site $i$. The energy degeneracy of $\Delta^{p_x\pm
ip_y}$ is lifted in the presence of magnetic field and the
$\Delta^{p_x+ip_y}$ is energetically favored when the field is
applied along $\hat{z}$, resulting in
$\Delta^{p_x+ip_y}({\bf{r}})=|\Delta(r)|e^{-i\phi}$ (winding
number -1).
% in consistence with the conclusion in the single
%vortex study of chiral $p_x\pm ip_y$-wave superconductor in a
%circular system \cite{matsumoto}.
Combining with our further identification of such behavior in the
$d_{x^2-y^2}\pm id_{xy}$-wave case where $d_{x^2-y^2}+id_{xy}$ is
favored, it indicates that the internal phase winding of the
Cooper pairs will try to counteract the phase winding of the
vortex to save the energy cost of supercurrents. To study the
vortex lattice structure of $\Delta^{p_x+ip_y}$, we select a
favorable electron occupancy $\bar{n}=1.2$ and $V=1.6$, giving
rise to the bulk value of $\Delta^{p_x+ip_y}=0.043$. Such a small
OP value results in a gap opened at $\Delta_{\text{Gap}}=0.14$
with a large core size according to the estimation $k_F\xi\sim
2E_F/\pi\Delta_{\text{Gap}}\simeq 27$. Figure \ref{p-fll} shows
the spatial distribution of the dominant
$\Delta^{p_x+ip_y}({\bf{r}})$ together with the induced
subdominant $\Delta^{p_x-ip_y}$ component, both with obvious
six-fold symmetry. Because the magnitude of $\Delta^{p_x+ip_y}$ is
small, it is sensitive to the magnetic field, resulting in a large
modulation of the magnitude of OP. The induced subdominant
$\Delta^{p_x-ip_y}$ is about one third of $\Delta^{p_x+ip_y}$. The
spatial structure of the subdominant $\Delta^{p_x-ip_y}$ has some
peculiar properties as displayed in Figs.~\ref{p-fll}(b1) and (b2)
which has not been shown before to our knowledge. We find that in
addition to the original vortices (OV) [small green disks in Fig.~
\ref{p-fll}(b2), with winding number +1], inter-vortex vortices
(IVV) [green triangles in Fig.~\ref{p-fll}(b2), with winding
number -1] are generated within every three OV. Therefore, each OV
is surrounded by six IVV and each IVV by three OV and three IVV.
The IVV forms honeycomb vortex lattice with length of the side
$\frac{1}{\sqrt{3}}$ of that of the OV lattice.
%However, the windings of
%the IVV are facing {\it frustration} from the viewpoint of the
%central OV and they try to form super-lattice structures as shown
%in Fig.~\ref{p-fll}(b2) where the group of vortices are enclosed
%by black and blue lines. The vortices on the black lines have
%winding number +1 while those on blue lines -1, leading to a
%honeycomb vortex lattice with length of the side
%$\frac{2}{\sqrt{3}}$ of that of the OV lattice. The small embedded
%honeycomb has a side length $\frac{1}{2}$ of the larger one.
%This phenomenon is not unique to the $p_x\pm ip_y$-wave pairing state
%but quite general in the condition that the vortex lattice is
%triangular and there is competing subdominant pairing state, and
%especially apparent when the magnitude of the dominant OP is small
%and the field is high.
A similar behavior has also be found for the $d_{x^2-y^2}\pm
id_{xy}$-wave case.
\begin{figure}
\includegraphics[width=7 cm]{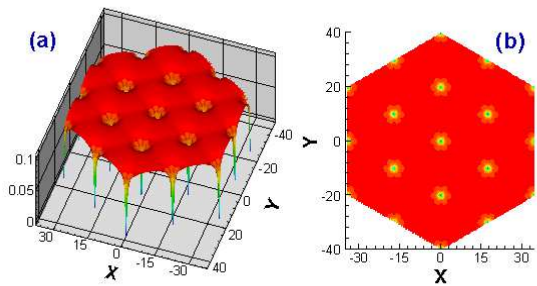}
\caption{3D (a) and contour plots (b) of the spatial distribution
of $|\Delta^{d_{x^2-y^2}+id_{xy}}|$. The MUC here is $20\times
40$.} \label{d-fll}
\end{figure}
For larger gap values, we study the vortex lattice structure of
$\Delta^{d_{x^2-y^2}\pm id_{xy}}$. We choose the electron
occupancy $\bar{n}=0.67$ and $V=1.3$, which leads to the bulk
value of $\Delta^{d_{x^2-y^2}+id_{xy}}=0.10$
($\Delta_{\text{Gap}}=0.4$). The spatial pattern of
$|\Delta^{d_{x^2-y^2}+id_{xy}}({\bf{r}})|$ is shown in Fig.~
\ref{d-fll}.
%Here, $V$ is smaller than that in the above studied
%$p_x+ip_y$-wave case but the OP is larger, which leading to vortex
%lattice with core size and spatial modulation small.
 The subdominant $\Delta^{d-id^\prime}$ (not shown here) has a similar
structure to $\Delta^{p_x-ip_y}$ [Fig.~\ref{p-fll}(b)] and the IVV
also forms honeycomb vortex lattice structures.
% although not as
% clear as in the small-gapped $\Delta^{p_x+ip_y}$ case.

\begin{figure}
\includegraphics[width=7 cm]{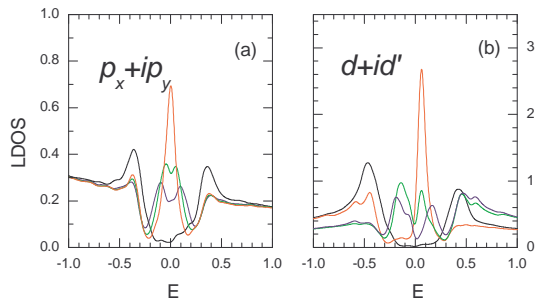}
\caption{The LDOS of $p_x+ip_y$- (a), $d_{x^2-y^2}+id_{xy}$-wave
(b) vortex lattice without ferromagnetic order (U=0) at the vortex
center (red lines), NN site of vortex center (green lines),  next
NN site of vortex center (blue lines) and midway of two nearest
vortices (black lines). $V=2$, $\bar{n}=1.2$ for (a); and $V=1.3$,
$\bar{n}=0.67$ for (b). The thermal broadening temperature is
$0.02$.} \label{ldos-no-fm}
\end{figure}
The local density of states (LDOS) as a function of energy of the
two pairing states are displayed in Fig.~\ref{ldos-no-fm} with
definition $\rho({\bf r}_i,E)=-\sum_{n}[|u_{i,\uparrow}^n|^2
f^{\prime}(E_n-E)+|v_{i,\downarrow}^n|^2 f^{\prime}(E_n+E)]$,
which is proportional to the differential tunnelling conductance
observed in scanning tunnelling microscopy (STM) experiments. For
clarity, we set $V=2.0$ with $\bar{n}=1.2$ to enlarge the
$p_x+ip_y$-wave OP ($\Delta^{p_x+ip_y}=0.09$) and $V=1.3$ with
$\bar{n}=0.67$ for the $d_{x^2-y^2}+id_{xy}$ pairing state so that
the gaps opened by them are comparable. Note that the LDOS of the
$p_x+ip_y$-wave pairing state is much lower than that of the
$d_{x^2-y^2}+id_{xy}$-wave case because the chemical potential is
sitting on the electronic spectrum position where the density of
states of the normal state is low. In the midway between two NN
vortices, the LDOS resembles that in bulk: both the chiral
$p_x+ip_y$-wave and $d_{x^2-y^2}+id_{xy}$-wave pairing states open
full gaps. Consequently, the low-lying quasiparticle bound states
emerge within the gap as expected, similar to the conventional
$s$-wave vortex core states\cite{caroli}. However, the energies of
the lowest core states in the vortex center of both the $p_x+ip_y$
and $d_{x^2-y^2}+id_{xy}$-wave pairing states deviate the
approximate relation $E_1=-\Delta^2_0/E_F$ (Note that $t<0$) for
conventional $s$-wave superconductors. $E_1$ of $p_x+ip_y$ vortex
is zero (pinned on the Fermi level) while that of
$d_{x^2-y^2}+id_{xy}$-wave vortex positive (above the Fermi
level). The difference of the bound state energy between
$p_x+ip_y$- and $d_{x^2-y^2}+id_{xy}$-wave vortex states is
nontrivial and is intrinsic to the internal angular momentum $l_z$
of the Cooper pairs. For the $p_x+ip_y$-wave state, the
quasiparticle wave functions $u$ and $v$ have 0 angular momentum
reflecting the total effect of the phase winding -1 of vortex and
$l_z=1$ of Cooper pairs, which accordingly gives rise to a bound
state with {\it strictly zero} energy \cite{matsumoto}. Similarly,
for the Cooper pair ($l_z=2$) with $d_{x^2-y^2}+id_{xy}$-wave
pairing symmetry\cite{franz}, $u$ has 0 and $v$ -1 angular
momentum thus with a positive bound energy\cite{matsumoto}. This
novel difference of vortex core bound states between the two
gapped chiral $p_x+ip_y$ and $d_{x^2-y^2}+id_{xy}$-wave pairing
state can be observed by STM experiments with high energy
resolution and might help to identify the pairing symmetry in this
material.

\begin{figure}
\includegraphics[width=7.5 cm]{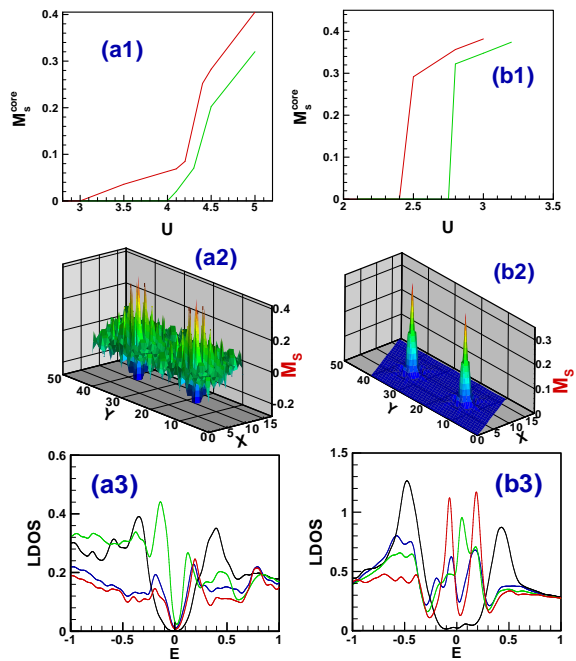}
\caption{(a1): The induced AFM moment at the $p_x+ip_y$-wave
vortex core $M_s^\text{core}$ as a function of $U$ for
$\bar{n}=1.2$ (red) and $\bar{n}=1.3$ (green). (a2): The spatial
structure of $M_s$ with $\bar{n}=1.2$ and $U=5$. (b1): The induced
FM moment at the $d_{x^2-y^2}+id_{xy}$-wave vortex core
$M_s^\text{core}$ as a function of $U$ for $\bar{n}=0.6$ (red) and
$\bar{n}=0.7$ (green). (b2): The spatial structure of $M_s$ with
$\bar{n}=0.7$ and $U=3$.
} %
\label{ldos-fm}
\end{figure}

We then study the induced magnetic moment around the vortex core
by examining the magnetization defined as
$M_s({\bf{r}}_i)=n_{i\uparrow}-n_{i\downarrow}$ and its dependence
on $U$ and $\bar{n}$. For the electron-doped case, we find that in
the presence of on-site repulsion the frustrated AFM moment might
be nucleated near the core for small doping as analogy to the case
of cuprates. The magnetic moment at the $p_x+ip_y$-wave vortex
core, $M_s^\text{core}$, as a function of $U$ with $\bar{n}=1.2$
and $\bar{n}=1.3$ is shown in Fig.~\ref{ldos-fm}(a1) with fixed
$V=2.0$. The critical value $U_\text{cr}$ increases while
$M_s^\text{core}$ decreases with $\bar{n}$ and we find no magnetic
moment for large doping with $\bar{n}=1.4$ up to $U=5$. And larger
$V$ also results in larger $U_\text{cr}$ because superconductivity
competes with magnetism. For the hole-doped region with
$\bar{n}<0.8$, a localized FM (instead of AFM) moment is induced
around the $d_{x^2-y^2}+id_{xy}$-wave vortices, completely
different from the picture of field-induced AFM order in
high-$T_c$ $d$-wave superconductors. Fig.~\ref{ldos-fm}(b1)
displays the $U$ dependence of $M_s^\text{core}$ for $\bar{n}=0.6$
and $\bar{n}=0.7$. Contrary to the electron doped case, larger
doping gives rise to weaker $U_\text{cr}$ and stronger magnetic
moment. We find that these seemingly surprising results have
little relevance to the pairing symmetry, but are intrinsic to the
competition between the AFM and FM orders in our model. The AFM
state dominates the region near the half filling, while the FM or
paramagnetic metallic states dominates the region near the Van
Hove singularity($0.5\lesssim\bar{n}<1$). The profiles of $M_s$
for the $p_x+ip_y$-wave vortex state with AFM moment and the
$d_{x^2-y^2}+id_{xy}$-wave case with the FM moment are displayed
in Fig.~\ref{ldos-fm}(a2),(b2). Fig.~\ref{ldos-fm}(a2) shows clear
staggered AFM manner of $M_s$ with slow decay while (b2) FM manner
with exponential decay away from the core. Different from the
charge density waves (CDW) with periodic modulations $4a$ in
$d$-wave cuprates\cite{tuvmodel}, we find the Friedel oscillation
of the electron density for the AFM case. As expected, both the
AFM and FM orders cause the double-peak splitting of the LDOS
peaks around the vortex center due to the lifting of the spin
up-down degeneracy as shown in Fig.~\ref{ldos-fm}(a3), (b3). Such
splitting of the LDOS associated with the vortex bound states
opens a symmetric ($p_x+ip_y$-wave case) or asymmetric
($d_{x^2-y^2}+id_{xy}$-wave case) subgap with respect to the Fermi
level, which provides a remarkable signal for the STM probing of
possible magnetic orderings in this material.

To summarize, we have investigated the novel vortex state of
Na$_x$CoO$_2$$\cdot y$H$_2$O with two possible pairing symmetries
realized in the 2D triangular lattice. Besides the intriguing
spatial structure of the vortex, we also find signature in the
electronic structure of the vortex state associated with different
pairing symmetries.
% especially the bound state with zero energy in
%the $p_x\pm ip_y$-wave vortex core and state with positive energy
%in the $d_{x^2-y^2}\pm id_{xy}$-wave case.
 In the presence of
strong on-site repulsion, we find frustrated AFM state in the
$p_x+ip_y$-wave vortex state and a localized FM state in the
$d_{x^2-y^2}\pm id_{xy}$-wave case.
%Double-peak structures of the
%LDOS near the Fermi level are found for both cases.
 The local
electronic structure and induced magnetic orders in the vortex
state might be observed by the microscopic STM\cite{pan} and
the spatially resolved NMR\cite{mitrovic} probes with high resolution.

 In finalizing this Letter, we noticed that Zhu and Balatsky
\cite{jxzhu2} addressed a similar issue with taking into account
only the $d_{x^2-y^2}\pm id_{xy}$-wave pairing. To our
understanding, at least the following  major considerations and
conclusions are quite different from theirs: (i)
%they only studied the spin-singlet $d_{x^2-y^2}+id_{xy}$-wave pairing while
we compared two superconducting states with different pairing
symmetries and found that when $\bar{n}=1.35$ [corresponding to
the $\bar{n}=0.65$ for $t>0$ in their paper] the ground state is
$p_x\pm ip_y$-wave pairing state, rather than  $d_{x^2-y^2}\pm
id_{xy}$ state; (ii) we found the AFM order induced in the core of
the $p_x+ip_y$-wave vortex ($U=5$) and localized FM order ($U=3$)
in the $d_{x^2-y^2}\pm id_{xy}$-wave vortex and predict that the
splitting of the LDOS peak by both of the magnetic orders; (iii)
we chose the {\it triangular} vortex lattice matching the
triangular CoO$_2$ lattice, with the vortex structure exhibiting
the intriguing six-fold symmetry, while they studied the {\it
square} vortex lattice.
%different from that of the real-space lattice in their paper.

We thank Y. Chen for useful discussions. This work was supported
by the RGC grant of Hong Kong(HKU7075/03P),
the 973-project of the Ministry of Science and
Technology of China under Grant Nos. G1999064602,
and the NSFC grants 10204011, 10204019, and 10021001.


\begin{thebibliography}{99}
\bibitem{comment1}
If $t>0$ $\bar{n}$ should be changed to $2-\bar{n}$.
 Correspondingly $p_x\pm ip_y$-wave pairing symmetry is
favored for $\bar{n}<1$ while $d_{x^2-y^2}\pm id_{xy}$-wave
pairing state for $1<\bar{n}<1.5$.
\end{thebibliography}
\end{document}